\documentclass{elsart}

\usepackage{amsfonts,amsmath}

\newcommand{\vect}[1]{\mbox{\boldmath $\mathbf{#1}$}}
\newcommand{\ie}{\emph{i.e.}}

\newcounter{bla}
\newenvironment{refnummer}{%
\list{[\arabic{bla}]}%
{\usecounter{bla}%
  \setlength{\itemindent}{0pt}%
  \setlength{\topsep}{0pt}%
  \setlength{\itemsep}{0pt}%
  \setlength{\labelsep}{2pt}%
  \setlength{\listparindent}{0pt}%
  \settowidth{\labelwidth}{[9]}%
  \setlength{\leftmargin}{\labelwidth}%
  \addtolength{\leftmargin}{\labelsep}%
  \setlength{\rightmargin}{0pt}}}
  {\endlist}

\begin{document}
\begin{frontmatter}

\title{Ground state of the time-independent Gross-Pitaevskii equation}

\author[a]{Claude M. Dion},
\ead{claude.dion@tp.umu.se}
\author[b]{Eric Canc\`{e}s}
\ead{cances@cermics.enpc.fr}


\address[a]{Department of Physics, Ume{\aa} University, SE-90187
   Ume{\aa}, Sweden}
\address[b]{CERMICS, \'{E}cole Nationale des Ponts et Chauss\'{e}es and
   INRIA, 6
   \& 8, avenue Blaise Pascal, cit\'{e} Descartes,
   F-77455 Marne-la-Vall\'{e}e Cedex 2, France}

\begin{abstract}
   We present a suite of programs to determine the ground state of the
   time-independent Gross-Pitaevskii equation, used in the simulation
   of Bose-Einstein condensates.  The calculation is based on the
   Optimal Damping Algorithm, ensuring a fast convergence to the true
   ground state. Versions are given for the one-, two-, and
   three-dimensional equation, using either a spectral method, well
   suited for harmonic trapping potentials, or a spatial grid.

\begin{flushleft}
PACS: 03.75.Hh; 03.65.Ge; 02.60.Pn; 02.70.-c

\end{flushleft}

\begin{keyword}
Gross-Pitaevskii equation; Bose-Einstein condensate; ground state;
Optimal Damping Algorithm.
\end{keyword}

\end{abstract}

\end{frontmatter}


{\bf PROGRAM SUMMARY}

\begin{small}
\noindent
{\em Manuscript Title:} Ground state of the time-independent
Gross-Pitaevskii equation  \\
{\em Authors:} Claude M. Dion and Eric Canc\`{e}s \\
{\em Program Title:}  GPODA \\
{\em Journal Reference:}                                      \\
{\em Catalogue identifier:}                                   \\
{\em Licensing provisions:} none                                  \\
{\em Programming language:} Fortran 90                          \\
{\em Computer:}  any                                           \\
{\em Compilers under which the program has been tested:} Absoft Pro
Fortran, The Portland Group Fortran 90/95 compiler, Intel Fortran  
Compiler\\
{\em RAM:} From $< 1$ MB in 1D to $\sim 10^2$ MB for a large 3D grid\\
{\em Keywords:} Gross-Pitaevskii equation, Bose-Einstein condensate,
Optimal Damping Algorithm \\
{\em PACS:}   03.75.Hh; 03.65.Ge; 02.60.Pn; 02.70.-c  \\
{\em Classification:}  2.7 Wave Functions and Integrals, 4.9
Minimization and Fitting  \\
{\em External routines:}   External FFT or eigenvector routines may be
required \\

{\em Nature of problem:} \\
The order parameter (or wave function) of a Bose-Einstein condensate
(BEC) is obtained, in a mean field approximation, by the
Gross-Pitaevskii equation (GPE)~[1].  The GPE is a nonlinear
Schr\"{o}dinger-like equation, including here a confining potential.   
The
stationary state of a BEC is obtained by finding the ground state of
the time-independent GPE, \ie, the order parameter that minimizes the
energy.  In addition to the standard three-dimensional GPE, tight
traps can lead to effective two- or even one-dimensional BECs, so the
2D and 1D GPEs are also considered.
\\
{\em Solution method:} \\
The ground state of the time-independent of the GPE is calculated
using the Optimal Damping Algorithm~[2].  Two sets of programs are
given, using either a spectral representation of the order
parameter~[3], suitable for a (quasi) harmonic trapping potential, or
by discretizing the order parameter on a spatial grid.
\\
{\em Running time:}\\
From seconds in 1D to a few hours for large 3D grids.
   \\
{\em References:}
\begin{refnummer}
\item F.~Dalfovo, S.~Giorgini, L.~P. Pitaevskii, S.~Stringari,
   Rev. Mod. Phys. 71 (1999) 463.
\item E.~Canc\`{e}s, C.~Le~Bris, Int. J. Quantum Chem. 79 (2000) 82.
\item C.~M. Dion, E.~Canc\`{e}s, Phys. Rev. E 67 (2003) 046706.
\end{refnummer}

\end{small}

\newpage


\hspace{1pc}
{\bf LONG WRITE-UP}

\section{Introduction}

Advances in cooling methods for dilute atomic gases have made it
possible to attain a new state of matter, the Bose-Einstein condensate
(BEC)~\cite{bec:pethick02,bec:pitaevskii03}.  As the temperature of
atoms gets very low, their de~Broglie wavelength, an inherently
quantum character, can become greater than the interatomic distance.
At that point, bosonic atoms will ``condense'' into a unique quantum
state and become indistinguishable parts a of macroscopic quantum
object, the BEC.  It has now been achieved for all stable alkali
atoms~\cite{bec:anderson95,bec:davis95,bec:bradley95,bec:cornish00,%
bec:weber03},
as well as with hydrogen~\cite{bec:fried98}, metastable
helium~\cite{bec:robert01,bec:pereira01}, and for diatomic
molecules~\cite{bec:jochim03}.

Starting from the many-body Hamiltonian describing the cold atoms, it
is possible to reduce the problem, by considering the order parameter,
or wave function, for the condensed fraction only.  It is governed by
a nonlinear Schr\"{o}dinger equation, the Gross-Pitaevskii equation
(GPE)~\cite{gp:gross61,gp:pitaevskii61,bec:dalfovo99,gp:stenholm98,gp:lieb00}
\begin{equation}
   \left[ - \frac{\hbar^{2}}{2m} \vect{\nabla}_{\vect{x}}^2 +
     V_{\mathrm{trap}}(\vect{x}) + \lambda_{\mathrm{3D}} \left| \psi 
(\vect{x})
     \right|^{2} \right] \psi(\vect{x}) = \mu \psi(\vect{x}) ,
   \label{eq:gp}
\end{equation}
with the normalization condition $\left\| \psi
\right\|_{L^{2}} = 1$, where $\hbar$ is the reduced Planck constant, $m$
the mass of the boson, $V_{\mathrm{trap}}$ a trapping
potential spatially confining the condensate, and $\mu$ the chemical
potential of the condensate.  Physically, the nonlinearity
corresponds to the mean field exerted on one boson by all the others
and is given, for a condensate of $N$ bosons in 3D, by
\begin{equation}
  \lambda_{\mathrm{3D}} \equiv g_{\mathrm{3D}} N = \frac{4 \pi \hbar^{2} a N}{m}.
  \label{eq:lambda}
\end{equation}
The value of $a$, the scattering length, varies according to the
species of bosons being considered.  The energy associated with the
wave function $\psi(\vect{x})$ is obtained according
to~\cite{gp:gross61,gp:pitaevskii61,bec:dalfovo99,gp:stenholm98,gp:lieb00}
\begin{equation}
  E [\psi] = N \int_{\mathbb{R}^3} \left[ \frac{\hbar^2}{2m} \left| \vect{\nabla}
      \psi(\vect{x}) \right|^2 + V_{\mathrm{trap}}(\vect{x}) \left|
      \psi(\vect{x}) \right|^{2} + \frac{\lambda_{\mathrm{3D}}}{2} \left|
      \psi(\vect{x}) \right|^{4} \right] \d\vect{x}.
\label{eq:energy}
\end{equation}

We present here a suite of programs designed to calculate the ground
state of the GPE, \ie, the order parameter $\psi(\vect{x})$ with to
the lowest energy.  This corresponds to the actual condensate order
parameter, in the absence of any excitation.  The problem is thus to
find the ground state of the condensate, that is a normalized
function $\psi_{\rm GS}(\vect{x})$ that minimizes $E[\psi]$.
Recall that if $V_{\mathrm{trap}}$ is continuous and goes to $+\infty$
at infinity, and if $\lambda_{\mathrm{3D}} \ge 0$, the ground state of
$E[\psi]$ exists and is unique
up to a global phase. In addition, the global phase can be chosen such
that $\psi_{\rm GS}$ is real-valued, and positive on $\mathbb{R}^3$.
The ground state $\psi_{\rm GS}$ can be computed using the Optimal
Damping Algorithm
(ODA), originally developed for
solving the Hartree-Fock equations~\cite{oda:cances00a,oda:cances00b}.
This algorithm is garanteed to
converge to the ground state.  Two different discretizations of
the order parameter are available in our sets of programs. In one case,
a basis set of eigenfunctions of the harmonic oscillator
is used, which is particularly suited for a harmonic (or
quasi-harmonic) trapping potential $V_{\mathrm{trap}}$.  In this case,
an efficient method to convert from the spectral representation to a
spatial grid~\cite{gp:dion03} is employed to treat the nonlinearity.
In the other case, a spatial grid is used throughout, with the kinetic
energy derivative evaluated with the help of Fast Fourier Transforms.
Note that, in all cases, the value of the energy given on output is
actually the energy per particle, $E[\psi]/N$.

\section{Optimal Damping Algorithm}

To describe the ODA~\cite{oda:cances00a,oda:cances00b} in the context
of the GPE, we start by defining the operators
\begin{equation}
   \hat{H}_0 \equiv - \frac{\hbar^{2}}{2m} \nabla_{\vect{x}}^2 +
   V(\vect{x}),
\end{equation}
corresponding to the linear part of the GPE~(\ref{eq:gp}), and
\begin{equation}
   \hat{H}(\rho) \equiv \hat{H}_0 + \lambda_{\mathrm{3D}} \rho(\vect{x})
\end{equation}
the full, nonlinear Hamiltonian, where we have introduced $\rho \equiv
\left| \psi \right|^2$ ($N \, \rho(\vect{x})$ is the density of the
condensate at point $\vect{x}$).

The ODA is based on the fact that the ground state density matrix
$\gamma_{\rm GS} = |\psi_{\rm GS}\rangle \langle \psi_{\rm GS}|$  
is the
unique minimizer of
\begin{equation} \label{eq:minODA}
\inf \left\{ {\cal E}[\gamma], \; \gamma \in {\cal
     S}(L^2(\mathbb{R}^3)), \; 0 \le \gamma \le I, \; \mbox{tr} 
(\gamma) = 1
\right\}.
\end{equation}
In the above minimization problem, ${\cal
     S}(L^2(\mathbb{R}^3))$ denotes the vector space of bounded
   self-adjoint operators on $L^2(\mathbb{R}^3)$ and $I$ the identity
   operator on~$L^2(\mathbb{R}^3)$. The energy functional ${\cal E} 
[\gamma]$ is
   defined by
$$
{\cal E}[\gamma] = \mbox{tr}(\hat{H}_0\gamma) +
\frac{\lambda_{\mathrm{3D}}}2 \int_{\mathbb{R}^3} \rho_\gamma^2,
$$
where $\rho_\gamma(\vect{x}) = \gamma(\vect{x},\vect{x})$
($\gamma(\vect{x},\vect{y})$ being the kernel of the trace-class
operator $\gamma$). The ODA implicitly generates a minimizing
sequence~$\gamma_k$ for
(\ref{eq:minODA}), starting, for instance, from the initial guess
$\gamma_0 =  |\psi_0 \rangle \langle \psi_0|$, where
$\psi_0$ is the ground state of~$\hat{H}_0$. The iterate $\gamma_{k+1}$
is constructed from the previous iterate $\gamma_k$ in two steps:
\begin{itemize}
\item Step~1: compute a normalized order parameter $\psi_k'$ which  
minimizes
$$
s_k = \inf \left\{ \left. \frac{\d}{\d t} {\cal E}\left[(1-t) \gamma_k + t |\psi
     \rangle \langle \psi| \right] \right|_{t=0}, \quad \| \psi \|_{L^2} = 1
\right\} .
$$
It is easy to check that $\psi_k'$ is in fact the ground state of
$\hat{H}(\rho_{\gamma_k})$ and that either $\psi_k' = \psi_{\rm GS}$ (up
to a global phase) or $s_k < 0$.
\item Step~2: compute
$$
\alpha_k = \mbox{arginf} \left\{  {\cal E}\left[(1-t) \gamma_k + t |\psi_k'
     \rangle \langle \psi_k'|\right], \quad t \in [0,1] \right\}
$$
and set $\gamma_{k+1} = (1-\alpha_k) \gamma_k + \alpha_k |\psi_k'
\rangle \langle \psi_k'|$. Note that $\alpha$ can be computed
analytically, for the function $t \mapsto {\cal E}\left[(1-t) \gamma_k
  + t |\psi_k' \rangle \langle \psi_k'|\right]$ is a second order
polynomial of the form ${\cal E}[\gamma_k] + t s_k + \frac{t^2}2 c_k$.
\end{itemize}
The set
$$
{\cal C} = \left\{ \gamma \in {\cal
     S}(L^2(\mathbb{R}^3)), \; 0 \le \gamma \le I, \; \mbox{tr} 
(\gamma) = 1
\right\}
$$
being convex, $\gamma_{k} \in {\cal C}$ for all $k$ and either
$\gamma_k = \gamma_{\rm GS}$ or ${\cal E}[\gamma_{k+1}] < {\cal
   E}[\gamma_k]$. In addition, it can be proved that, up to a global
phase, $\psi_k'$ converges to $\psi_{\rm GS}$ when $k$ goes to
infinity. Likewise, $\rho_k \equiv \rho_{\gamma_k}$ converges to
$\rho_{\rm GS} \equiv \psi_{\rm GS}^2$. It is important to note that the
sequences $\psi_k'$ and $\rho_k$ can be generated without
explicitely computing $\gamma_k$. This is crucial to reduce the overall
memory requirement of~ODA.

Let us now describe a practical implementation of ODA, in which only
order parameters and densities are stored in memory.
The algorithm is initialized by $\psi_0$, from which we derive
$\rho_0 = \left| \psi_0 \right|^2$, $f_0 = (\psi_0,\hat{H}_0 \psi_0) 
$, and $h_0
= (\psi_0,\hat{H}(\rho_0) \psi_0)$.  The iterations go as follows:
\begin{enumerate}
\item Calculate the ground state $\psi_{k}'$ of $\hat{H}(\rho_k)$, and
  $\rho_{k}' = \left| \psi_k' \right|^2$.

\item Compute
  \begin{eqnarray}
    f'_{k} & = & (\psi_{k}',\hat{H}_0 \psi_{k}'), \nonumber \\
    h'_{k} & = & (\psi_{k}',\hat{H}(\rho_k) \psi_{k}'), \nonumber \\
    h''_{k} & = & (\psi_{k}',\hat{H}(\rho_{k}') \psi_{k}'). \nonumber
  \end{eqnarray}

\item Calculate
  \begin{eqnarray}
    s_k & = & h_{k}' - h_{k}, \nonumber \\
    c_k & = & h_{k} + h_{k}'' - 2 h_{k}' + f_{k}' - f_{k}. \nonumber
  \end{eqnarray}

\item Set $\alpha_k = 1$ if $c_k \leq -s_k$, $\alpha_k = -s_k/c_k$
  otherwise, and
  \begin{eqnarray}
    E_{\mathrm{opt}} & = & \frac{1}{2} \left( f_{k} + h_{k} \right) +
    \alpha_k s_k + \frac{\alpha_k^2}{2} c_k, \nonumber \\
    \rho_{k+1} & = & (1-\alpha_k) \rho_{k} + \alpha_k \rho_{k}',  
    \nonumber \\
    f_{k+1} & = &  (1-\alpha_k) f_{k} + \alpha_k f_{k}',
    \nonumber \\
    h_{k+1} & = & 2 E_{\mathrm{opt}} - f_{k+1}. \nonumber
  \end{eqnarray}

\item If $|s_k/E_{\mathrm{opt}}| > \varepsilon_{\mathrm{ODA}}$
  (convergence criterion), go to (1), otherwise compute the ground
  state of $H(\rho_{k+1})$, which is the solution sought, and
  terminate.
\end{enumerate}

To calculate the ground state of the operators $\hat{H}_0$ and
$\hat{H}(\rho)$, the inverse power method is used, with the
convergence criterion $\left| E_{i+1} - E_{i} \right| \leq
\varepsilon_{\mathrm{IP}}$, where $E$ are the lowest eigenvalues at
consecutive iterations.  The inverse power algorithm itself uses the
conjugated gradient method to solve $\hat{H} v = u$, with $u$ given
and $v$ unknown.  The convergence of the conjugated gradient is
controlled by the criterion $\varepsilon_{\mathrm{CG}}$.  The only
exception to this is in \texttt{gpoda1Ds}, where the ground states of
the operators are found by a matrix eigenproblem solver routine (see
Sec.~\ref{sec:usr1Ds}).

\section{Representations of the GPE}

The Gross-Pitaevskii equation was defined in Eq.~(\ref{eq:gp}), with
the nonlinearity Eq.~(\ref{eq:lambda}) in 3D.  In this work, we are
also considering cases where the confinement $V_{\mathrm{trap}}$ is so
tight in some spatial dimension that the condensate can actually be
considered as a two-, or even one-dimensional object.  This leads to
different representations of the nonlinearity $\lambda$ and the
expression for the coupling parameters $g_{\mathrm{2D}}$ and
$g_{\mathrm{1D}}$ can be found in
Refs.~\cite{bec:olshanii98,gp:petrov00,gp:lee02}.  We refer to chapter
17 of~\cite{bec:pitaevskii03} for a detailed discussion of the
validity of the mean field approximation in these cases.

\subsection{Spatial grid approach}
\label{sec:grid}

If the order parameter is represented on a discretized spatial grid,
the calculation of the potential energy and the nonlinearity are
trivial, as they both act locally, while the kinetic energy operator
is non-local.  By means of a Fourier transform, it is possible to
convert from position to momentum space, where the kinetic operator is
local.  This is implemented by means of a Fast Fourier Transforms
(FFTs), allowing to convert back and forth between the two
representations, to evaluate each part of the Hamiltonian in the space
where it is local.

\subsection{Spectral method}

For many situations, the trapping potential is harmonic, or a close
variation thereof, \ie,
\begin{equation}
   V_{\mathrm{trap}}(x,y,z) = \frac{m}{2} \left( \omega_x^2 x^2 +
     \omega_y^2 y^2  +  \omega_z^2 z^2  \right) + V_0(x,y,z),
\label{eq:potential}
\end{equation}
where $\omega$ is the trapping frequencies in each direction and $V_0$
accounts for eventual corrections to a purely harmonic trap.  In this
case, it is advantageous to use a basis set made up of the
eigenfunctions of the quantum harmonic oscillator.

We start by rescaling Eq.~(\ref{eq:gp}), introducing dimensionless
lengths $(\tilde{x},\tilde{y},\tilde{z})$,
\begin{subequations}
   \begin{equation}
     x =  \left( \frac{\hbar}{m \omega_x} \right)^{1/2} \tilde{x},
   \end{equation}
   \begin{equation}
     y =  \left( \frac{\hbar}{m \omega_y} \right)^{1/2} \tilde{y},
   \end{equation}
   \begin{equation}
     z = \left( \frac{\hbar}{m \omega_z} \right)^{1/2} \tilde{z},
   \end{equation}
\label{eq:rescale}
\end{subequations}
and a new order parameter $\tilde{\psi}$ defined as
\begin{equation}
\psi(x,y,z) = A\tilde{\psi}(x,y,z).
\end{equation}
Considering the normalization condition
\begin{equation}
\int_{\mathbb{R}^3} \left| \psi(x,y,z) \right|^2 \d x \d y \d z = 1,
\end{equation}
we take
\begin{equation}
   A = \left( \frac{m}{\hbar} \right)^{3/4} \left( \omega_x \omega_y
     \omega_z \right)^{1/4}
\end{equation}
such that
\begin{equation}
\int_{\mathbb{R}^3} \left| \tilde{\psi}(\tilde{x},\tilde{y},\tilde{z})
\right|^2 \d \tilde{x} \d \tilde{y} \d \tilde{z} = 1.
\end{equation}
The Gross-Pitaevskii equation now reads
\begin{multline}
   \biggl[ \frac{\omega_x}{\omega_z} \left( - \frac{1}{2}\nabla^{2}_{\tilde 
{x}}
       + \frac{\tilde{x}^2}{2} \right) + \frac{\omega_y}{\omega_z}  
\left(
       - \frac{1}{2} \nabla^{2}_{\tilde{y}} + \frac{\tilde{y}^2}{2}  
\right) +
     \left( - \frac{1}{2} \nabla^{2}_{\tilde{z}} + \frac{\tilde{z}^2}{2}
     \right) \\
     + \tilde{V}_0(\tilde{x}, \tilde{y}, \tilde{z}) +
   \tilde{\lambda}_{\mathrm{3D}} \left|
       \tilde{\psi}(\tilde{x},\tilde{y},\tilde{z})
   \right|^{2} \biggr] = \tilde{\mu}
\tilde{\psi}(\tilde{x},\tilde{y},\tilde{z}),
\label{eq:gptilde}
\end{multline}
with
\begin{equation}
   \tilde{V}_0(\tilde{x}, \tilde{y}, \tilde{z}) \equiv \frac{1}{\hbar
     \omega_z} V_0(x,y,z),
\label{eq:V0tilde}
\end{equation}
\begin{equation}
   \tilde{\lambda}_{\mathrm{3D}} \equiv \frac{m^{3/2}}{\hbar^{5/2}}  
\left(
     \frac{\omega_x \omega_y}{\omega_z} \right)^{1/2} g_{\mathrm{3D}}  
N =
   4 \pi a N \left( \frac{m}{\hbar} \frac{\omega_x \omega_y}{\omega_z}
     \right)^{1/2},
\label{eq:lambdatilde3D}
\end{equation}
and
\begin{equation}
\tilde{\mu} \equiv \frac{\mu}{\hbar \omega_z}.
\label{eq:mutilde}
\end{equation}
Similarly,
\begin{equation}
\tilde{E}[\tilde{\psi}] \equiv \frac{E[\psi]}{\hbar \omega_z}.
\label{eq:Etilde}
\end{equation}

Using the Galerkin approximation, we can express the order parameter
$\tilde{\psi}$ as a linear combination of a finite number of
(orthonormal) basis functions $\phi$,
\begin{equation}
   \tilde{\psi}(\tilde{x}, \tilde{y}, \tilde{z}) = \sum_{i = 0}^{N_ 
{\tilde{x}}}
   \sum_{j = 0}^{N_{\tilde{y}}}  \sum_{k = 0}^{N_{\tilde{z}}} c_{ijk}
   \phi_i(\tilde{x})  \phi_j(\tilde{y})  \phi_k(\tilde{z}),
\label{eq:spectr}
\end{equation}
where the $\phi$ are chosen as the eigenfunctions of the 1D harmonic
oscillator, \ie,
\begin{equation}
   \left( - \frac{1}{2} \frac{\d^2}{\d\xi^2} + \frac{\xi^2}{2} \right)  
\phi_n(\xi)
   = \left( n + \frac{1}{2} \right) \phi_n(\xi).
\label{eq:ho}
\end{equation}
In the spectral representation of Eq.~(\ref{eq:spectr}),
Eq.~(\ref{eq:gptilde}) becomes a series of coupled equations for the
coefficients $c_{ijk}$, and the first part of the Hamiltonian can be
evaluated by a simple multiplication, according to Eq.~(\ref{eq:ho}).
The second part of the Hamiltonian, consisting of the $\tilde{V}_0$
and the nonlinear terms, is local in $(\tilde{x}, \tilde{y},
\tilde{z})$ and couples the different coefficients.  Its operation
can be calculated in a manner similar to what is used for the spatial
grid (see Sec.~\ref{sec:grid}): starting from the coefficients
$c_{ijk}$, the order parameter $\tilde{\psi}$ is evaluated at selected
grid points $(\tilde{x}, \tilde{y}, \tilde{z})$, the local terms are
then trivially calculated, and the order parameter is transformed back
to the spectral representation.  This procedure can be performed
efficiently and accurately using the method described in
Ref.~\cite{gp:dion03}.

For the 2D case, \ie, when the motion along $y$ is suppressed, we
rescale the lengths according to Eq.~(\ref{eq:rescale}), which results
in
\begin{equation}
   A = \left( \frac{m}{\hbar} \right)^{1/2} \left( \omega_x
     \omega_z \right)^{1/4}
\end{equation}
for the scaling factor of the order parameter.  We thus obtain the
2D GPE
\begin{equation}
   \left[\frac{\omega_x}{\omega_z} \left( - \frac{1}{2}\nabla^{2}_{\tilde 
{x}}
       + \frac{\tilde{x}^2}{2} \right) + \left( - \frac{1}{2}
       \nabla^{2}_{\tilde{z}} + \frac{\tilde{z}^2}{2} \right)
     + \tilde{V}_0(\tilde{x}, \tilde{z}) +
     \tilde{\lambda}_{\mathrm{2D}} \left|
       \tilde{\psi}(\tilde{x},\tilde{z}) \right|^{2} \right] =
   \tilde{\mu} \tilde{\psi}(\tilde{x},\tilde{z}),
\label{eq:gptilde2D}
\end{equation}
where
\begin{equation}
   \tilde{\lambda}_{\mathrm{2D}} \equiv \lambda_{\mathrm{2D}}
   \frac{m}{\hbar^2} \left( \frac{\omega_x}{\omega_z} \right)^{1/2}.
\label{eq:lambdatilde2D}
\end{equation}

Similarly, we get for the one-dimensional case (where the motion along
$x$ and $y$ is frozen)
\begin{equation}
   A = \left( \frac{m \omega_z}{\hbar} \right)^{1/4},
\end{equation}
\begin{equation}
   \left[  - \frac{1}{2}
       \nabla^{2}_{\tilde{z}} + \frac{\tilde{z}^2}{2}
     + \tilde{V}_0(\tilde{z}) +
     \tilde{\lambda}_{\mathrm{1D}} \left|
       \tilde{\psi}(\tilde{z}) \right|^{2} \right] =
   \tilde{\mu} \tilde{\psi}(\tilde{z}),
\label{eq:gptilde1D}
\end{equation}
and
\begin{equation}
   \tilde{\lambda}_{\mathrm{1D}} \equiv \lambda_{\mathrm{1D}} \left(
   \frac{m}{\hbar^3 \omega_z} \right)^{1/2}.
\label{eq:lambdatilde1D}
\end{equation}

\section{Description of the programs}

\subsection{\textrm{\tt gpoda3Dg}}

This program solves the full 3D GPE~(\ref{eq:gp}) on a grid.  Atomic
units are used throughout.

\subsubsection{User-supplied routines}
\label{sec:3Dg_user}

The double precision function \texttt{potentialV(x,y,z)} takes as
input the three double precision arguments \texttt{x}, \texttt{y}, and
\texttt{z}, corresponding to the spatial coordinates $(x,y,z)$, and
returns $V_{\mathrm{trap}}(x,y,z)$.

A 3D FFT routine must also be supplied.  The program is set up to work
with the \textsc{dfftpack}~\cite{fftpack:swarztrauber82}
transform of a real function, and can be linked directly to this
library.

If the user wishes to use another FFT, the file \texttt{fourier3D.f90}
must be modified accordingly.  The program first calls
\texttt{fft\_init(n)}, where \texttt{n} is a one-dimensional integer
array of length 4, the last three elements containing the number of
grid points in $x$, $y$, and $z$, with the first element corresponding
to the maximum number of grid points in any direction, \ie, for
\texttt{n(0:3)}, \texttt{n(0) = maxval(n(1:3))}.  The program will
then call repeatedly the subroutine
\texttt{fourier3D(n,fin,fout,direction)}, with \texttt{fin} and
\texttt{fout} double precision arrays of dimension
\texttt{(n(1),n(2),n(3))}, and \texttt{direction} an integer.  The
routine should return in array \texttt{fout} the forward Fourier
transform of \texttt{fin} if $\texttt{direction} = 1$, and the inverse
transform for $\texttt{direction} = -1$. Any variable initialized by
\texttt{fft\_init} must be passed to \texttt{fourier3D} through a
module.  Note that the main program expects to receive the Fourier
coefficients (following the forward transform) according to:
\begin{eqnarray}
   c_1 & = & \sum_{n=1}^{N} f_n, \nonumber \\
   c_{2m-2} & = & \sum_{n=1}^{N} f_n \cos \left[ \frac{2\pi (m-1)
       (n-1)}{N} \right], \ \ m = 2,\ldots,N/2+1 \nonumber \\
   c_{2m-1} & = & -\sum_{n=1}^{N} f_n \sin \left[ \frac{2\pi (m-1)
       (n-1)}{N} \right], \ \ m = 2,\ldots,N/2 \nonumber
\end{eqnarray}
where the coefficients $c_m$ correspond to variable \texttt{fout} and
the sequence $f_n$ to \texttt{fin}.

\subsubsection{Input parameters}
\label{sec:input3Dg}

The input parameters are read from a namelist contained in a file
named \texttt{params3Dg.in}, with the following format (the variable
type is indicated in parenthesis, where dp stands for double precision):
\begin{tabbing}
\texttt{\&}\=\texttt{params3Dg} \\
  \> \texttt{mass =} \emph{mass of the boson (dp)}\texttt{,} \\
  \> \texttt{lambda =} \emph{nonlinearity $\lambda_{\mathrm{3D}}$
    (dp)}\texttt{,} \\
  \> \texttt{ng\_x =} \emph{number of grid points in $x$,
    (integer)}\texttt{,} \\
  \> \texttt{ng\_y =} \emph{number of grid points in $y$,
    (integer)}\texttt{,} \\
  \> \texttt{ng\_z =} \emph{number of grid points in $z$,
    (integer)}\texttt{,} \\
  \> \texttt{xmin =} \emph{first point of the grid in $x$ (dp)}\texttt 
{,} \\
  \> \texttt{xmax =} \emph{last point of the grid in $x$ (dp)}\texttt 
{,} \\
  \> \texttt{ymin =} \emph{first point of the grid in $y$ (dp)}\texttt 
{,} \\
  \> \texttt{ymax =} \emph{last point of the grid in $y$ (dp)}\texttt 
{,} \\
  \> \texttt{zmin =} \emph{first point of the grid in $z$ (dp)}\texttt 
{,} \\
  \> \texttt{zmax =} \emph{last point of the grid in $z$ (dp)}\texttt 
{,} \\
  \> \texttt{critODA =} \emph{convergence criterion for the ODA,
    $\varepsilon_{\mathrm{ODA}}$ (dp)}\texttt{,} \\
  \> \texttt{critIP =} \emph{convergence criterion for the inverse  
power,
    $\varepsilon_{\mathrm{IP}}$ (dp)}\texttt{,}  \\
  \> \texttt{critCG =} \emph{convergence criterion for the conjugated  
gradient,
    $\varepsilon_{\mathrm{CG}}$ (dp)}\texttt{,}  \\
  \> \texttt{itMax =} \emph{maximum number of iterations of the ODA
    (integer)}\texttt{,} \\
  \> \texttt{guess\_from\_file =} \emph{read initial guess from file}
  \texttt{guess3Dg.data}\emph{? (logical)}  \\
\texttt{\&end}
\end{tabbing}

If the value of the input parameter \texttt{guess\_from\_file} is
\texttt{.true.}, a file named \texttt{guess3Dg.data} must be present in
the local directory.
It contains the initial guess for the order parameter, and must
consist in \texttt{ng\_x} $\times$ \texttt{ng\_y} $\times$
\texttt{ng\_z} lines, each containing the values of the coordinates
$x$, $y$, and $z$, followed by $\psi(x,y,z)$.  \emph{Note that the
   program does \textbf{not} check if the coordinates correspond to the
   grid defined by the input parameters.}  The program will simply
assign the first value of $\psi$ to the first grid point,
$(x_{\mathrm{min}}, y_{\mathrm{min}}, z_{\mathrm{min}})$, then the
second value to the second grid point in $x$, with $y =
y_{\mathrm{min}}$ and $z = z_{\mathrm{min}}$, etc.  After $n_x$ points
have been read, the next value of $\psi$ is assigned to the second
grid point in $y$, with $x = x_{\mathrm{min}}$ and $z =
z_{\mathrm{min}}$, and so on.  In other words, the fourth column of
\texttt{guess3Dg.data} contains $\psi(x,y,z)$ in standard Fortran
format, with $x$ corresponding to the first index, $y$ to the second,
and $z$ to the third.

\subsubsection{Output files}

The order parameter is written out in file \texttt{gs3Dg.data}, with
each line containing the coordinates $x$, $y$, and $z$, followed by
$\psi(x,y,z)$.  If the algorithm has not converged, the file will
contain the function obtained at the last iteration. The format of
\texttt{gs3Dg.data} is the same as that of \texttt{guess3Dg.data} (see
Sec.~\ref{sec:input3Dg}), such that \texttt{gs3Dg.data} can be used as
an initial guess for a new run, with for instance a different value of
$\lambda$ (if the grid is changed, the function must be interpolated
to the new grid beforehand).

\subsection{\textrm{\tt gpoda2Dg}}

This program solves the 2D GPE on a grid, corresponding to the 3D case
where motion along $y$ is frozen.  Atomic units are used throughout.

\subsubsection{User-supplied routines}

The double precision function \texttt{potentialV(x,z)} takes as
input the two double precision arguments \texttt{x} and \texttt{z},
corresponding to the spatial coordinates $(x,z)$, and returns
$V_{\mathrm{trap}}(x,z)$.

A 2D FFT routine must also be supplied.  The program is set up to work
with the \textsc{dfftpack}~\cite{fftpack:swarztrauber82}
transform of a real function, and can be linked directly to this
library.  For use of another FFT routine, please see
Sec.~\ref{sec:3Dg_user}.

\subsubsection{Input parameters}
\label{sec:input2Dg}

The input parameters are read from a namelist contained in a file
named \texttt{params2Dg.in}.  The namelist \texttt{\&params2Dg}
follows the same format as the namelist \texttt{\&params3Dg} presented
in Sec.~\ref{sec:input3Dg}, with the omission of variables
\texttt{ng\_y}, \texttt{ymin}, and \texttt{ymax}.  Also, the parameter
\texttt{lambda} corresponds here to $g_{\mathrm{2D}} N$
\cite{gp:petrov00,gp:lee02}.

If the value of the input parameter \texttt{guess\_from\_file} is
\texttt{.true.}, a file named \texttt{guess2Dg.data} must be present in
the local directory.
The format of the file is similar to that of \texttt{guess3Dg.data},
presented in Sec.~\ref{sec:input3Dg}, with the exception of data
corresponding to coordinate $y$.

\subsubsection{Output files}

The order parameter is written out in file \texttt{gs2Dg.data}, with
each line containing the coordinates $x$ and $z$, followed by
$\psi(x,z)$.  If the algorithm has not converged, the file will
contain the function obtained at the last iteration.  The format of
\texttt{gs2Dg.data} is the same as that of \texttt{guess2Dg.data} (see
Sec.~\ref{sec:input2Dg}), such that \texttt{gs2Dg.data} can be used as
an initial guess for a new run, with for instance a different value of
$\lambda_{\mathrm{2D}}$ (if the grid is changed, the function must be
interpolated to the new grid beforehand).

\subsection{\textrm{\tt gpoda1Dg}}

This program solves the 1D GPE on a grid, corresponding to the 3D case
where motion along $x$ and $y$ is frozen.  Atomic units are used
throughout.

\subsubsection{User-supplied routines}

The double precision function \texttt{potentialV(z)} takes as
input the double precision argument \texttt{z}, corresponding to the
spatial coordinate $z$, and returns $V_{\mathrm{trap}}(z)$.

An FFT routine must also be supplied.  The program is set up to work
with the \textsc{dfftpack}~\cite{fftpack:swarztrauber82}
transform of a real function, and can be linked directly to this
library.  For use of another FFT routine, please see
Sec.~\ref{sec:3Dg_user}.

\subsubsection{Input parameters}
\label{sec:input1Dg}

The input parameters are read from a namelist contained in a file
named \texttt{params1Dg.in}.  The namelist \texttt{\&params1Dg}
follows the same format as the namelist \texttt{\&params3Dg} presented
in Sec.~\ref{sec:input1Dg}, with the omission of variables
\texttt{ng\_x}, \texttt{ng\_y}, \texttt{xmin}, \texttt{xmax},
\texttt{ymin}, and \texttt{ymax}.  Also, the parameter \texttt{lambda}
corresponds here to $g_{\mathrm{1D}} N$~\cite{bec:olshanii98}.

If the value of the input parameter \texttt{guess\_from\_file} is
\texttt{.true.}, a file named \texttt{guess1Dg.data} must be present in
the local directory.
It contains the initial guess for the order parameter, and must
consist in \texttt{ng\_z} lines, each containing the values of the
coordinate $z$ followed by $\psi(z)$.  \emph{Note that the program
   does \textbf{not} check if the coordinates correspond to the grid
   defined by the input parameters.}  The program will simply assign
the first value of $\psi$ to the first grid point, $z_{\mathrm{min}}$,
then the second value to the second grid point in $z$, and so on.

\subsubsection{Output files}

The order parameter is written out in file \texttt{gs1Dg.data}, with
each line containing the coordinate $z$ followed by $\psi(z)$.  If the
algorithm has not converged, the file will contain the function
obtained at the last iteration.  The format of \texttt{gs1Dg.data} is
the same as that of \texttt{guess1Dg.data} (see
Sec.~\ref{sec:input1Dg}), such that \texttt{gs1Dg.data} can be used as
an initial guess for a new run, with for instance a different value of
$\lambda_{\mathrm{1D}}$ (if the grid is changed, the function must be
interpolated to the new grid beforehand).

\subsection{\textrm{\tt gpoda3Ds}}

This program solves the full 3D GPE~(\ref{eq:gptilde}) using a
spectral method.  Note that the value of \texttt{mu} calculated is
actually the rescaled $\tilde{\mu}$ defined by Eq.~(\ref{eq:mutilde}).

\subsubsection{User-supplied routines}

The double precision function \texttt{potentialV0(x,y,z)} takes
as input the three double precision arguments \texttt{x}, \texttt{y},
and \texttt{z}, corresponding to the rescaled spatial coordinates
$(\tilde{x},\tilde{y},\tilde{z})$, and returns $\tilde{V}_0(\tilde{x},
\tilde{y}, \tilde{z})$, defined by Eq.~(\ref{eq:V0tilde}).

\subsubsection{Input parameters}
\label{sec:input3Ds}

The input parameters are read from a namelist contained in a file
named \texttt{params3Ds.in}, with the following format (the variable
type is indicated in parenthesis, where dp stands for double
precision):
\begin{tabbing}
\texttt{\&}\=\texttt{params3Ds} \\
  \> \texttt{lambda =} \emph{nonlinearity
    $\tilde{\lambda}_{\mathrm{3D}}$ [Eq.~(\ref{eq:lambdatilde3D})]
    (dp)}\texttt{,} \\
  \> \texttt{wxwz =} \emph{trap frequency ratio $\omega_x/\omega_z$
    (dp)}\texttt{,} \\
  \> \texttt{wywz =} \emph{trap frequency ratio $\omega_y/\omega_z$
    (dp)}\texttt{,} \\
  \> \texttt{n\_x =} \emph{highest basis function in $x$, $N_{\tilde 
{x}}$
    (integer)}\texttt{,} \\
  \> \texttt{n\_y =} \emph{highest basis function in $y$, $N_{\tilde 
{y}}$
    (integer)}\texttt{,} \\
  \> \texttt{n\_z =} \emph{highest basis function in $z$, $N_{\tilde 
{z}}$
    (integer)}\texttt{,} \\
  \> \texttt{symmetric\_x =} \emph{symmetric potential in $x$
    (logical)}\texttt{,} \\
  \> \texttt{symmetric\_y =} \emph{symmetric potential in $y$
    (logical)}\texttt{,} \\
  \> \texttt{symmetric\_z =} \emph{symmetric potential in $z$
    (logical)}\texttt{,} \\
  \> \texttt{critODA} = \emph{convergence criterion for the ODA,
    $\varepsilon_{\mathrm{ODA}}$ (dp)}\texttt{,} \\
  \> \texttt{critIP} = \emph{convergence criterion for the inverse  
power,
    $\varepsilon_{\mathrm{IP}}$ (dp)}\texttt{,}  \\
  \> \texttt{critCG} = \emph{convergence criterion for the conjugated  
gradient,
    $\varepsilon_{\mathrm{CG}}$ (dp)}\texttt{,}  \\
  \> \texttt{itMax} = \emph{maximum number of iterations of the ODA
    (integer)}\texttt{,} \\
  \> \texttt{guess\_from\_file} = \emph{read initial guess from file}
  \texttt{guess3Ds.data}\emph{? (logical)}  \\
  \> \texttt{output\_grid =} \emph{write final order parameter to file}
  \texttt{gs3Ds\_grid.data}\emph{? (logical)}  \\
\texttt{\&end}
\end{tabbing}

The algorithm used to find the roots of the Hermite polynomial, needed
for the spectral method~\cite{gp:dion03}, limits the acceptable
highest basis function to $\texttt{n} \leq 91$.  The value of the
parameters \texttt{symmetric} allow to reduce the size of the basis
set used, for the case where the additional trapping potential $V_0$
[Eq.~(\ref{eq:potential})] is even along any of the axes.  For
instance, if $V_0(x,y,z) = V_0(-x,y,z)$, setting \texttt{symmetric\_x
   = .true.} will restrict the basis set along $x$ to even functions
$\phi(x)$ [Eq.~(\ref{eq:spectr})], as the order parameter will present
the same parity as the trapping potential $V_{\mathrm{trap}}$.  Note
that in all cases the parameters \texttt{n} set the index of the
highest harmonic oscillator eigenfunction used, not the number of
basis functions used.

If the value of the input parameter \texttt{guess\_from\_file} is
\texttt{.true.}, a file named \texttt{guess3Ds.data} must be present in
the local directory.
It contains the initial guess for the order parameter and contains
lines with the values of indices $i$, $j$, and $k$ (all integers),
followed by the coefficient $c_{ijk}$ (double precision), see
Eq.~(\ref{eq:spectr}).  If an index is greater than the value of $N$
for the corresponding spatial axis, or if its parity is not consistent
with the chosen symmetry (see above), it is ignored.  If a set of
indices $ijk$ appears more than once, only the last value of $c_{ijk}$
is kept, and any $c_{ijk}$ not specified in the file is taken to be
equal to zero.

If the value of the input parameter \texttt{output\_grid} is
\texttt{.true.}, a second namelist will be read from the file
\texttt{params3Ds.in}:
\begin{tabbing}
\texttt{\&}\=\texttt{grid3D} \\
  \> \texttt{ng\_x =} \emph{number of grid points in $\tilde{x}$,
    (integer)}\texttt{,} \\
  \> \texttt{ng\_y =} \emph{number of grid points in $\tilde{y}$,
    (integer)}\texttt{,} \\
  \> \texttt{ng\_z =} \emph{number of grid points in $\tilde{z}$,
    (integer)}\texttt{,} \\
  \> \texttt{xmin =} \emph{first point of the grid in $\tilde{x}$
    (dp)}\texttt{,} \\
  \> \texttt{xmax =} \emph{last point of the grid in $\tilde{x}$
    (dp)}\texttt{,} \\
  \> \texttt{ymin =} \emph{first point of the grid in $\tilde{y}$
    (dp)}\texttt{,} \\
  \> \texttt{ymax =} \emph{last point of the grid in $\tilde{y}$
    (dp)}\texttt{,} \\
  \> \texttt{zmin =} \emph{first point of the grid in $\tilde{z}$
    (dp)}\texttt{,} \\
  \> \texttt{zmax =} \emph{last point of the grid in $\tilde{z}$ (dp)}
  \\
\texttt{\&end}
\end{tabbing}
(see next section for details on usage).

\subsubsection{Output files}
\label{sec:output3Ds}

The order parameter is written out in file \texttt{gs3Ds.data}, with
each line containing the indices $i$, $j$, and $k$, followed by the
coefficients $c_{ijk}$ of Eq.~(\ref{eq:spectr}).  If the algorithm has
not converged, the file will contain the function obtained at the last
iteration. The format of \texttt{gs3Ds.data} is the same as that of
\texttt{guess3Ds.data} (see Sec.~\ref{sec:input3Ds}), such that
\texttt{gs3Ds.data} can be used as an initial guess for a new run,
with for instance a different value of $\tilde{\lambda}$.

If the value of the input parameter \texttt{output\_grid} is
\texttt{.true.}, the order parameter is also written out to the file
\texttt{gs3Ds\_grid.data}, with each line containing the coordinates
$\tilde{x}$, $\tilde{y}$, and $\tilde{z}$, defined by the namelist
\texttt{\&grid3D}, followed by
$\tilde{\psi}(\tilde{x},\tilde{y},\tilde{z})$.

\subsection{\textrm{\tt gpoda2Ds}}

This program solves the a 2D GPE using a spectral method.  Note
that the value of \texttt{mu} calculated is actually the rescaled
$\tilde{\mu}$ defined by Eq.~(\ref{eq:mutilde}).

\subsubsection{User-supplied routines}

The double precision function \texttt{potentialV0(x,z)} takes
as input the three double precision arguments \texttt{x} and
\texttt{z}, corresponding to the rescaled spatial coordinates
$(\tilde{x},\tilde{z})$, and returns
$\tilde{V}_0(\tilde{x},\tilde{z})$, defined by the 2D equivalent of
Eq.~(\ref{eq:V0tilde}).

\subsubsection{Input parameters}
\label{sec:input2Ds}

The input parameters are read from a namelist contained in a file
named \texttt{params2Ds.in}.  The namelist \texttt{\&params2Ds}
follows the same format as the namelist \texttt{\&params3Ds} presented
in Sec.~\ref{sec:input3Ds}, with the omission of variables
\texttt{wywz}, \texttt{n\_y}, and \texttt{symmetry\_y}.  Also, the
parameter \texttt{lambda} corresponds here to
$\tilde{\lambda}_{\mathrm{2D}}$ [Eq.~(\ref{eq:lambdatilde2D})].

If the value of the input parameter \texttt{guess\_from\_file} is
\texttt{.true.}, a file named \texttt{guess2Ds.data} must be present in
the local directory.
The format is the same as the file \texttt{guess3Ds.data}
(Sec.~\ref{sec:input3Ds}), except that only indices $i$ and $k$ are
present.

If the value of the input parameter \texttt{output\_grid} is
\texttt{.true.}, a second namelist named \texttt{\&grid2D} will be read
from the file \texttt{params2Ds.in}.  This namelist is the same as
\texttt{\&grid3D} of Sec.~\ref{sec:input3Ds}, without the variables
corresponding to $\tilde{y}$.

\subsubsection{Output files}
\label{sec:output2Ds}

The order parameter is written out in file \texttt{gs2Ds.data}, with
a format similar to file \texttt{gs3Ds.data} described in
Sec.~\ref{sec:output3Ds}, except that only indices $i$ and $k$ are
present.  If the value of the input parameter \texttt{output\_grid} is
\texttt{.true.}, the order parameter is also written out to the file
\texttt{gs2Ds\_grid.data}, in the same manner as for file
\texttt{gs3Ds\_grid.data} (Sec.~\ref{sec:output3Ds}), but without the
$\tilde{y}$ coordinate.

\subsection{\textrm{\tt gpoda1Ds}}

This program solves the a 1D GPE using a spectral method.  Note
that the value of \texttt{mu} calculated is actually the rescaled
$\tilde{\mu}$ defined by Eq.~(\ref{eq:mutilde}).

\subsubsection{User-supplied routines}
\label{sec:usr1Ds}

The double precision function \texttt{potentialV0(z)} takes as
input the three double precision arguments \texttt{z}, corresponding
to the rescaled spatial coordinate $\tilde{z}$, and returns
$\tilde{V}_0(\tilde{z})$, defined by the 1D equivalent of
Eq.~(\ref{eq:V0tilde}).

A routine for calculating eigenvalues and eigenvectors must be
supplied.  The program is set up to use the
\textsc{lapack}~\cite{lapack:anderson99} routine for the
eigenvalue problem for a real symmetric matrix.  To use another
routine, file \texttt{eigen1D.f90} has to be modified.  The subroutine
\texttt{eigen(n,H,eigenval,eigenvec)} takes as input the integer
\texttt{n} and the double precision array \texttt{H(n,n)}.  On output,
the double precision real \texttt{eigenval} and the double precision
array \texttt{eigenvec(n)} contain repectively the smallest eigenvalue
of matrix \texttt{H} and the corresponding eigenvector.

\subsubsection{Input parameters}
\label{sec:input1Ds}

The input parameters are read from a namelist contained in a file
named \texttt{params1Ds.in}, with the following format (the variable
type is indicated in parenthesis, where dp stands for double
precision):
\begin{tabbing}
\texttt{\&}\=\texttt{params1Ds} \\
  \> \texttt{lambda =} \emph{nonlinearity
    $\tilde{\lambda}_{\mathrm{1D}}$ [Eq.~(\ref{eq:lambdatilde1D})]
    (dp)}\texttt{,} \\
  \> \texttt{n =} \emph{highest basis function, $N$
    (integer)}\texttt{,} \\
  \> \texttt{symmetric =} \emph{spatially symmetric potential?
    (logical)}\texttt{,} \\
  \> \texttt{critODA} = \emph{convergence criterion for the ODA,
    $\varepsilon_{\mathrm{ODA}}$ (dp)}\texttt{,} \\
  \> \texttt{itMax} = \emph{maximum number of iterations of the ODA
    (integer)}\texttt{,} \\
  \> \texttt{guess\_from\_file} = \emph{read initial guess from file}
  \texttt{guess1Ds.data}\emph{? (logical)}  \\
  \> \texttt{output\_grid =} \emph{write final order parameter to file}
  \texttt{gs1Ds\_grid.data}\emph{? (logical)}  \\
\texttt{\&end}
\end{tabbing}

See Sec.~\ref{sec:input3Ds} for restrictions on the value of
\texttt{n} and the use of \texttt{symmetric}.

If the value of the input parameter \texttt{guess\_from\_file} is
\texttt{.true.}, a file named \texttt{guess1Ds.data} must be present in
the local directory.
The format is the same as the file \texttt{guess1Ds.data}
(Sec.~\ref{sec:input3Ds}), except that only index $k$ is present.

If the value of the input parameter \texttt{output\_grid} is
\texttt{.true.}, a second namelist named \texttt{\&grid1D} will be read
from the file \texttt{params1Ds.in}.  This namelist is the same as
\texttt{\&grid1D} of Sec.~\ref{sec:input3Ds}, without the variables
corresponding to $\tilde{x}$ and $\tilde{y}$.

\subsubsection{Output files}

The order parameter is written out in file \texttt{gs1Ds.data}, with a
format similar to file \texttt{gs1Ds.data} described in
Sec.~\ref{sec:output3Ds}, except that only index $k$ is present.  If
the value of the input parameter \texttt{output\_grid} is
\texttt{.true.}, the order parameter is also written out to the file
\texttt{gs1Ds\_grid.data}, in the same manner as for file
\texttt{gs1Ds\_grid.data} (Sec.~\ref{sec:output3Ds}), but without the
$\tilde{x}$ and $\tilde{y}$ coordinates.

\section*{Acknowledgments}

This research was conducted in part using the resources of the High
Performance Computing Center North (HPC2N).



\newpage
\hspace{1pc}
{\bf TEST RUN OUTPUT}

\bigskip

Considering a condensate of $10^4$ $^{87}$Rb atoms, in a harmonic trap
of frequency $\omega_x = \omega_y = \omega_z/\sqrt{8} = 2\pi \times 90\
\mathrm{Hz}$ with the parameter file \texttt{params3Ds.in} as follows:
\begin{verbatim}
&params3Ds
  lambda = 368.8d0,
  wxwz = 0.353553390593d0,
  wywz = 0.353553390593d0,
  n_x = 20,
  n_y = 20,
  n_z = 20,
  symmetric_x = .true.,
  symmetric_y = .true.,
  symmetric_z = .true.,
  critODA = 1.d-8,
  critIP = 1.d-8,
  critCG = 1.d-8,
  itMax = 100,
  guess_from_file = .false.,
  output_grid = .false.
&end
\end{verbatim}
the output will look like:
\begin{verbatim}
  GPODA3Ds

Parameters:

  omega_x / omega_z =   0.35355339E+00
  omega_y / omega_z =   0.35355339E+00
  Nonlinearity =        0.36880000E+03

  Number of basis functions:   11 x   11 x   11 =     1331
  Number of grid points:       41 x   41 x   41 =    68921
  Symmetric in x y z

  Initialization

    Compute the ground state of H_0
    Inverse Power converged in      2 iterations
     --> mu =  0.853553390593000


  Iteration   1

    Compute the ground state of H(psi_in)
    Inverse Power converged in     27 iterations
     --> mu =  2.19942774785621
    Optimal damping
      slope  -22.0705785783271
      step    0.768246751736393
      Eopt    4.08395470599574
\end{verbatim}
[...]
\begin{verbatim}
 Iteration   66
 
   Compute the ground state of H(psi_in)
   Inverse Power converged in      2 iterations
    --> mu =  3.90057925938285
   Optimal damping
     slope  -1.667024296381214E-008
     step    3.537833144766566E-002
     Eopt    2.87515659549269
 
 
 Convergence achieved in  66  iterations
  --> mu =  3.90057925938285
  --> E =   2.87515659549269
 
 
 Checking self-consistency
 
   Inverse Power converged in      2 iterations
    --> mu =  3.90057337542902
   l1 norm of psi2out-psi2in:   0.171325E-01 for   68921 grid points
   (0.248582E-06 per grid point)
\end{verbatim}
(Running time: 14 min on a 2.5 GHz PowerPC G5 Quad.)

\end{document}